\documentstyle[amssymb,epsfig,12pt]{article}

\input{tcilatex}

\begin{document}

\begin{titlepage}
\bigskip \begin{flushright}
\end{flushright}


\vspace{1cm}

\begin{center}
{\Large \bf {Kerr-AdS Bubble Spacetimes and Time-Dependent  AdS/CFT Correspondence}}\\
\end{center}
\vspace{2cm}
\begin{center}
 A.M. Ghezelbash{%
\footnote{%
EMail: amasoud@sciborg.uwaterloo.ca}} and R. B. Mann
\footnote{
EMail: mann@avatar.uwaterloo.ca}\\
Department of Physics, University of Waterloo, \\
Waterloo, Ontario N2L 3G1, Canada\\
\vspace{1cm}
PACS numbers:  
04.50.+h,11.10.Kk,04.20.Jb\\
Keywords: Time-dependent AdS/CFT correspondence,  Trace anomaly,  Bubble spacetime

\end{center}
\begin{abstract}
We compute the boundary stress-energies of time-dependent asymptotically AdS 
spacetimes in 5 and 7 dimensions, and find that their traces are equal 
to the respective 4 and 6 dimensional field-theoretic trace anomalies. This provides 
good supporting evidence in favour of the AdS/CFT correspondence in
time-dependent backgrounds. 
\end{abstract}
\end{titlepage}\onecolumn


\section{Introduction}

An essential ingredient in constructing a theory of quantum gravity is to
understand its behaviour in time-dependent settings. A recent promising
approach to this end is to construct simple time-dependent solutions that
provide (at least to leading order) consistent time-dependent backgrounds
for string theory. This has been carried out in asymptotically flat
spacetimes, and recently extended to include the asymptotically anti de
Sitter (AdS) case \cite{AHA,Bir,bala,CAI,MANN}. \ This latter situation is
of interest since the AdS/CFT correspondence conjecture could be employed to
relate the time-dependence to the behaviour of the non-perturbative field
theory dual. \ In the context of string theory this has recently led to
intensive investigation of asymptotically de Sitter spacetimes to see what
the prospects are for developing a de Sitter/CFT correspondence \cite{dsCFT}.

In this paper, we study in more detail the higher dimensional bubble
spacetimes derived from analytic continuation of \ odd-dimensional Kerr-AdS
spacetimes \cite{Hawk, awad1,awad2}. We find that the relationship between
the field-theoretic trace anomaly \cite{Henning} and the asymptotic boundary
stress-energy is in agreement with the AdS/CFT prediction, demonstrating
evidence for the correspondence for time-dependent backgrounds that are
asymptotically AdS.

Although the relationship between the boundary stress-energy tensor and the
trace anomaly have been previously studied \cite{Henning,deharo}, these
results cannot be directly applied to the time-dependent case due to
subtleties in these settings. As an example of these subtleties, in ref. %
\cite{cv} the authors studied the time-dependent metric obtained by analytic
continuation of the non-abelian T-dual of the Schwarzschild-AdS black hole.
The result of the calculation shows that the trace of the boundary
stress-energy tensor is different from the well known general trace anomaly.
To solve this discrepancy, the authors considered the dualized model in the
time direction and found that time-dependent AdS/CFT holds only on a special
subspace of the whole spacetime. These observations indicate that some
specific modifications may be needed in the application of \ the AdS/CFT
correspondence to time-dependent backgrounds.

Motivated by the above, in this paper we study the AdS/CFT correspondence
applied to time-dependent bubble spacetimes in five and seven dimensions.
Seven is the smallest dimensionality with acceptable bubble structure (no
signature changes or ergoregions \cite{MANN}) in which the correspondence
can be non-trivially tested. We compute the boundary stress-energies
associated for such seven-dimensional bubble solutions and show that their
traces are proportional to the Euler densities of the six-dimensional field
theory, in accord with the AdS/CFT correspondence for a time-dependent
setting. \ We also consider a double analytic continuation of
five-dimensional Kerr-AdS spacetime. Although this bubble spacetime has an
ergoregion (raising difficult issues as to its interpretation as a
background) we again find an equivalence between the trace of the boundary
stress-energy tensor and four-dimensional Euler density. The equivalence
shows that the AdS/CFT correspondence conjecture holds even though the
background metric changes sign. We extend our considerations to double
analytic continuation of five-dimensional Kerr-AdS spacetime with two
rotation parameters, again finding an equivalence between the trace of the
stress-energy tensor and four-dimensional Euler density despite the presence
of an ergoregion. Higher dimensional Kerr-AdS spacetimes have two acceptable
double analytic continuations \cite{Bir}. One involves the continuation of
one of the coordinates in the $d\Omega _{d-4}$ part of the $d$-dimensional
Kerr-AdS metric, yielding 
\begin{equation}
\begin{array}{c}
ds^{2}=\frac{\Delta _{r}}{\rho ^{2}}\{\ell d\chi +\frac{\alpha }{\Xi }%
(1-X^{2})d\phi \}^{2}+(1-X^{2})\frac{\Delta _{X}}{\rho ^{2}}\{a\ell d\chi -%
\frac{r^{2}\ell ^{2}-a^{2}}{\Xi }d\phi \}^{2} \\ 
+\frac{\rho ^{2}\ell ^{2}}{\Delta _{r}}dr^{2}+\frac{\rho ^{2}}{%
(1-X^{2})\Delta _{X}}dX^{2}+\ell ^{2}r^{2}X^{2}\{-d\tau ^{2}+\cosh ^{2}\tau
d\Omega _{d-5}^{2}\}%
\end{array}
\label{Kerr7bb1}
\end{equation}%
where 
\begin{equation}
\begin{array}{c}
\Delta _{r}=(r^{2}\ell ^{2}-a^{2})(1+r^{2})-\frac{2GM}{(r\ell )^{d-5}} \\ 
\Delta _{X}=1+\frac{a^{2}X^{2}}{\ell ^{2}}\text{ \ \ \ , \ \ \ \ }\Xi =1+%
\frac{a^{2}}{\ell ^{2}} \\ 
\rho ^{2}=r^{2}\ell ^{2}-a^{2}X^{2}%
\end{array}
\label{delta}
\end{equation}%
The bubble metric (\ref{Kerr7bb1}) has a time dependent $\left( d-4\right) $%
-dimensional de Sitter space provided $d\geq 6$. Since only a coordinate of
the $S^{d-4}$ is continued there is no Milne phase in the bubble evolution.

The other bubble metric can be obtained from continuation of the one angular
coordinate of the spherical section embedded in a $\left( d-4\right) $%
-dimensional hyperbolic space. The bubble metric is 
\begin{equation}
\begin{array}{c}
ds^{2}=\frac{\Delta _{r}}{\rho ^{2}}\{\ell d\chi -\frac{\alpha }{\Xi }%
(X^{2}-1)d\phi \}^{2}+(X^{2}-1)\frac{\Delta _{X}}{\rho ^{2}}\{a\ell d\chi -%
\frac{r^{2}\ell ^{2}-a^{2}}{\Xi }d\phi \}^{2} \\ 
+\frac{\rho ^{2}\ell ^{2}}{\Delta _{r}}dr^{2}+\frac{\rho ^{2}}{%
(X^{2}-1)\Delta _{X}}dX^{2}+\ell ^{2}r^{2}X^{2}\{d\psi ^{2}+\sinh ^{2}\psi
(-d\tau ^{2}+\cosh ^{2}\tau d\Omega _{d-6}^{2})\}%
\end{array}
\label{Kerr7bb2}
\end{equation}%
where 
\begin{equation}
\begin{array}{c}
\Delta _{r}=(r^{2}\ell ^{2}-a^{2})(r^{2}-1)-\frac{2GM}{(r\ell )^{d-5}} \\ 
\Delta _{X}=1-\frac{a^{2}X^{2}}{\ell ^{2}}\text{ \ \ \ , \ \ \ \ }\Xi =1-%
\frac{a^{2}}{\ell ^{2}}%
\end{array}
\label{delta2}
\end{equation}%
The bubble metric (\ref{Kerr7bb2}) again has a time dependent $\left(
d-4\right) $-dimensional de Sitter space, in this case provided $d\geq 7$.
As with the metric (\ref{Kerr7bb1}) there is no Milne phase because only an
angular coordinate is analytically continued.

\section{Boundary stress-energy tensor and trace anomaly for 7-dimensional
Kerr-AdS bubbles}

We compute the boundary stress-energy tensor for the bubble solutions (\ref%
{Kerr7bb1}) and (\ref{Kerr7bb2}) for $d=7$ using the well-known counterterm
method \cite{counterterm}. The results are 
\begin{equation}
\begin{array}{c}
8\pi GT_{\tau }^{\tau }=\frac{-1}{80\ell ^{7}r^{6}}(\pm 5\ell ^{6}-80\ell
^{2}GM\pm 46a^{4}X^{2}\ell ^{2}+51a^{6}X^{4}+3a^{6}X^{2}+3a^{2}\ell ^{4}X^{2}
\\ 
-55a^{6}X^{6}\mp 51a^{4}X^{4}\ell ^{2}+5a^{2}\ell ^{4}\mp 5a^{4}\ell
^{2}-5a^{6})+O(\frac{1}{r^{8}}) \\ 
8\pi GT_{\phi }^{\phi }=\frac{-1}{80\ell ^{7}r^{6}}(\pm 5\ell ^{6}\mp
101a^{4}\ell ^{2}-a^{2}\ell ^{4}-189a^{6}X^{2}+231a^{6}X^{4}-55a^{6}X^{6} \\ 
-80\ell ^{2}GM\pm 168a^{4}X^{2}\ell ^{2}+3a^{2}\ell ^{4}X^{2}\mp
51a^{4}X^{4}\ell ^{2}+25a^{6})+O(\frac{1}{r^{8}}) \\ 
8\pi GT_{X}^{X}=\frac{-1}{80\ell ^{7}r^{6}}(\pm 5\ell ^{6}-5a^{6}+5a^{2}\ell
^{4}\mp 5a^{4}\ell ^{2}-80\ell ^{2}GM\pm 66a^{4}X^{2}\ell ^{2}-3a^{2}\ell
^{4}X^{2} \\ 
+45a^{6}X^{4}-25a^{6}X^{6}-3a^{6}X^{2}\mp 45a^{4}X^{4}\ell ^{2})+O(\frac{1}{%
r^{8}}) \\ 
8\pi GT_{\chi }^{\chi }=\frac{1}{80\ell ^{7}r^{6}}(\pm 25\ell
^{6}+5a^{6}-101a^{2}\ell ^{4}\mp a^{4}\ell ^{2}-400\ell ^{2}GM\mp
168a^{4}X^{2}\ell ^{2} \\ 
+189a^{2}\ell ^{4}X^{2}-51a^{6}X^{4}+55a^{6}X^{6}-3a^{6}X^{2}\pm
231a^{4}X^{4}\ell ^{2})+O(\frac{1}{r^{8}}) \\ 
8\pi GT_{\chi }^{\phi }=\frac{3}{40\ell ^{8}r^{6}}a(-\ell ^{2}\mp
a^{2})(30a^{4}X^{4}\pm 32a^{2}\ell ^{2}X^{2}-32a^{4}X^{2}\mp 22a^{2}\ell
^{2}+5\ell ^{4}+5a^{4})+O(\frac{1}{r^{8}}) \\ 
8\pi GT_{\lambda }^{\lambda }=8\pi GT_{\psi }^{\psi }=8\pi GT_{\tau }^{\tau }%
\end{array}
\label{Kerr7bb1and2stress}
\end{equation}%
where the upper signs refer to the bubble (\ref{Kerr7bb1}) and the lower
signs to (\ref{Kerr7bb2}). From this relations, after rescaling with an
appropriate power of conformal factor, (the boundary metric diverges due to
an infinite conformal factor and so by rescaling of the boundary metric, the
CFT lives on a boundary without any divergences), we obtain 
\begin{equation}
\widetilde{T}_{\mu }^{\mu }=\frac{-3a^{2}}{32\pi G\ell ^{7}}%
(8a^{4}X^{4}-5a^{4}X^{6}+2\ell ^{4}\mp 2a^{2}\ell ^{2}\mp 8a^{2}X^{4}\ell
^{2}\pm 9a^{2}\ell ^{2}X^{2}-3a^{4}X^{2}-3\ell ^{4}X^{2})+O(\frac{1}{r^{2}})
\label{trace}
\end{equation}%
for the respective traces of the boundary stress-energies.

Alternatively the trace anomaly in the stress-energy tensor of a classically
Weyl-invariant even-dimensional quantum field{\bf \ }theory has the
structure \cite{Henning,Tseytlin} 
\begin{equation}
{\cal A=}A+B+D  \label{ABD}
\end{equation}%
where $A$ is proportional to the even-dimensional Euler density, $B$ is a
sum of independent Weyl invariants that contain the Weyl tensor and its
derivatives, and $D$ is a \ total derivative term of a covariant expression
that can be removed by adding suitable local counterterms. Since the
six-dimensional boundary metrics of the bubble spaces (\ref{Kerr7bb1}) and (%
\ref{Kerr7bb2}) are conformally flat, the Weyl invariant terms vanish. The
only non-vanishing term in (\ref{ABD}) is then the Euler density, which in
six dimensions is given by \cite{Henning}, 
\begin{equation}
E_{6}=\frac{1}{6912}%
(K_{1}-12K_{2}+3K_{3}+16K_{4}-24K_{5}-24K_{6}+4K_{7}+8K_{8})  \label{Esix}
\end{equation}%
in terms of metric invariants 
\begin{equation}
\begin{array}{c}
K_{1}={\sf R}^{3},K_{2}={\sf RR}_{\alpha \beta }{\sf R}^{\alpha \beta
},K_{3}={\sf RR}_{\alpha \beta \gamma \delta }{\sf R}^{\alpha \beta \gamma
\delta },K_{4}={\sf R}_{\alpha }^{\quad \beta }{\sf R}_{\beta }^{\quad
\gamma }{\sf R}_{\gamma }^{\quad \alpha },K_{5}={\sf R}^{\alpha \beta }{\sf R%
}^{\gamma \delta }{\sf R}_{\alpha \gamma \delta \beta } \\ 
K_{6}={\sf R}_{\alpha \beta }{\sf R}^{\alpha \gamma \delta \epsilon }{\sf R}%
_{\quad \gamma \delta \epsilon }^{\beta },K_{7}={\sf R}_{\alpha \beta \gamma
\delta }{\sf R}^{\alpha \beta \epsilon \zeta }{\sf R}_{\quad \epsilon \zeta
}^{\gamma \delta },K_{8}={\sf R}_{\alpha \beta \gamma \delta }{\sf R}%
^{\alpha \zeta \eta \delta }{\sf R}_{\quad \zeta \eta }^{\beta \quad \quad
\gamma }%
\end{array}
\label{K18}
\end{equation}%
where ${\sf R}_{\alpha \beta \gamma \delta }$ is the curvature tensor of the
boundary metric. Inserting (\ref{K18}) into (\ref{Esix}) we find that for
rescaled boundary metric, after a lengthy calculation%
\begin{equation}
E_{6}=\frac{a^{2}}{24\ell ^{12}}(-8a^{4}X^{4}+5a^{4}X^{6}-2\ell ^{4}\pm
2a^{2}\ell ^{2}\pm 8a^{2}X^{4}\ell ^{2}\mp 9a^{2}\ell
^{2}X^{2}+3a^{4}X^{2}+3\ell ^{4}X^{2})+O(\frac{1}{r^{2}})  \label{Esixkerr}
\end{equation}
for the respective bubbles (\ref{Kerr7bb1}) and (\ref{Kerr7bb2}) in the
large-$r$ limit.

Comparing our result (\ref{trace}) to the Euler density (\ref{Esixkerr}) of
the boundary theory we find 
\begin{equation}
\widetilde{T}_{\mu }^{\mu }=\alpha E_{6}  \label{traceE6}
\end{equation}%
where $\alpha =\frac{9\ell ^{5}}{4\pi G}=\frac{12N^{3}}{\pi ^{3}}.$ In the
last equation, we employed the relation $G=\frac{3\pi ^{2}\ell ^{5}}{16}%
\frac{1}{N^{3}}$ for the seven-dimensional Newton's constant. Note that our
definition of the Euler density (\ref{Esix}) is different by a factor of \ $%
\frac{-1}{55296}$\ with the definition of the Euler density considered in %
\cite{Tseytlin} which we denote it by $\widehat{E}_{6}$. In other words, $%
E_{6}=\frac{-1}{55296}\widehat{E}_{6}$, and so from the relation (\ref%
{traceE6}), we have%
\begin{equation}
\widetilde{T}_{\mu }^{\mu }=-\frac{N^{3}}{4608\pi ^{3}}\widehat{E}_{6}=\frac{%
4N^{3}}{(4\pi )^{3}7!}\{\frac{-35}{2}\widehat{E}_{6})  \label{traceE6hat}
\end{equation}%
which is in good agreement with what is predicted by AdS/CFT correspondence.
The conformal anomaly of the free 6-dimensional chiral (2,0) tensor
multiplet has been calculated in \cite{Tseytlin} and the result is similar
to (\ref{traceE6hat}), up to a well known factor of 4/7. The reason of this
difference is due to the fact that the coefficient in front of the Euler
density in six dimensions is related to the 4-point function which is not
governed by a non-renormalization theorem.

We recall here some points about the field-theoretic calculation of the
trace anomaly in the time-dependent metrics. The trace anomaly is related to
the logarithmically divergent part of the effective action $\Gamma =\frac{1}{%
2}\log \det \Delta $\ of any Euclidean field theory defined over a $d-$%
dimensional compact Riemannian manifold $M,$\ where $\Delta $\ is the
Laplace operator. After using the Seely-DeWitt asymptotic expansion \cite{DW}%
, the logarithmically divergent part of the effective action takes the form\ 
\begin{equation}
\Gamma _{\infty }=-\frac{1}{2}\log \frac{L^{2}}{\mu ^{2}}\int_{M}d^{d}x\sqrt{%
g}b_{d}  \label{gammainf}
\end{equation}%
where $L$ is an UV cut-off and then the trace anomaly $T_{\mu }^{\mu }$\ is
simply equal to $b_{d}$\ , the Seely coefficient \cite{Tseytlin} of the
non-singular term in the Seely-DeWitt asymptotic expansion of the operator $%
\Delta $.

The above field theoretic trace anomaly calculation is valid if we switch to
a time-independent Lorentzian field theory by performing a Wick rotation on
Euclidean field theory. For time-dependent\ Lorentzian field theory (which
at least\thinspace $\ $locally could change to a static form by some Wick
rotations), the trace anomaly calculation also is valid since in this case,
after applying the proper Wick rotations, we get an Euclidean field theory.
\ As an example, in the next section, the boundary Euclidean field theory
for a bubble spacetime after Wick rotation is considered.

The calculation of $b_{d}$\ shows that the general form of the trace anomaly
for an even $d$-dimensional field theory is in the form of (\ref{ABD}).

\section{Boundary stress-energy tensor and trace anomaly for 5-dimensional
Kerr-AdS bubbles with one and two rotational parameters}

We now extend these considerations to Kerr-AdS bubbles in five dimensions.
The time dependent five-dimensional metric is given by

\begin{equation}
\begin{array}{c}
ds^{2}=\frac{\Delta _{r}}{\rho ^{2}}\{d\chi +\frac{\alpha }{\Xi }%
(1+X^{2})d\phi \}^{2}+(1+X^{2})\frac{\Delta _{X}}{\rho ^{2}}\{ad\chi -\frac{%
r^{2}-a^{2}}{\Xi }d\phi \}^{2} \\ 
+\frac{\rho ^{2}}{\Delta _{r}}dr^{2}-\frac{\rho ^{2}}{(1+X^{2})\Delta _{X}}%
dX^{2}+r^{2}X^{2}d\lambda ^{2}%
\end{array}
\label{Kerr5bb}
\end{equation}%
where 
\begin{equation}
\begin{array}{c}
\Delta _{r}=(r^{2}-a^{2})(1+r^{2}/\ell ^{2})-2GM \\ 
\Delta _{X}=1+\frac{a^{2}X^{2}}{\ell ^{2}}\text{ \ \ \ , \ \ \ \ }\Xi =1+%
\frac{a^{2}}{\ell ^{2}} \\ 
\rho ^{2}=r^{2}+a^{2}X^{2}%
\end{array}
\label{delta3}
\end{equation}%
These bubbles have ergoregions, unlike their seven-dimensional counterparts.
For (\ref{Kerr5bb}) the condition 
\begin{equation}
\Delta _{X}<0  \label{ergoreg}
\end{equation}%
defines the ergoregion.

Hence (\ref{Kerr5bb}) describes a well defined bubble metric provided $%
\left| X\right| <X_{cr}$ where $X_{cr}=\ell /a.$ While this might render
them unsuitable as pure bubble metrics \cite{Bir}, we find that trace
anomaly in of the boundary stress-energy tensor (\ref{ABD}) is related to
the four-dimensional Euler density of the boundary metric of the bubble (\ref%
{Kerr5bb}), in accord with the AdS/CFT prediction. Specifically we find that
the large-$r$ boundary stress-energy tensor components for the bubble
spacetime (\ref{Kerr5bb}) are 
\begin{equation}
\begin{array}{c}
8\pi GT_{\phi }^{\phi }=-\frac{1}{8\ell r^{4}}\{(7X^{2}+10)X^{2}a^{4}-2a^{2}%
\ell ^{2}(X^{2}+1)-\ell ^{2}(\ell ^{2}+8GM)+3a^{4}\}+O\left( \frac{1}{r^{6}}%
\right) \\ 
8\pi GT_{X}^{X}=-\frac{1}{8\ell r^{4}}\{(3X^{2}+2)X^{2}a^{4}-2a^{2}\ell
^{2}(X^{2}+1)-\ell ^{2}(\ell ^{2}+8GM)-a^{4}\}+O\left( \frac{1}{r^{6}}\right)
\\ 
8\pi GT_{\chi }^{\chi }=-\frac{1}{8\ell r^{4}}\{(7X^{2}+2)X^{2}a^{4}-2a^{2}%
\ell ^{2}(5X^{2}+1)+3\ell ^{2}(\ell ^{2}+8GM)-a^{4}\}+O\left( \frac{1}{r^{6}}%
\right) \\ 
8\pi GT_{\lambda }^{\lambda }=-\frac{1}{8\ell r^{4}}%
\{(7X^{2}+2)X^{2}a^{4}-2a^{2}\ell ^{2}(X^{2}+1)-\ell ^{2}(\ell
^{2}+8GM)-a^{4}\}+O\left( \frac{1}{r^{6}}\right) \\ 
8\pi GT_{\chi }^{\phi }=-\frac{1}{2\ell ^{3}r^{4}}a(\ell
^{2}+a^{2})(2a^{2}X^{2}-\ell ^{2}+a^{2})+O\left( \frac{1}{r^{6}}\right)%
\end{array}
\label{Kerr5bbstress}
\end{equation}%
and so its trace after rescaling is 
\begin{equation}
\widetilde{T}_{\mu }^{\mu }=-\frac{a^{2}}{8\pi G\ell }%
\{a^{2}X^{2}(3X^{2}+2)-\ell ^{2}(1+2X^{2})\}+O\left( \frac{1}{r^{2}}\right)
\label{trace2}
\end{equation}%
The trace anomaly (\ref{ABD}) of the four-dimensional boundary of \ the
bubble spacetime has the same form of (\ref{ABD}), where $A$ is proportional
to the Euler density in four dimensions. The $B$ term vanishes since the
boundary metric is conformally flat. The Euler density is given by the
relation \cite{Henning} 
\begin{equation}
E_{4}=\frac{1}{64}(\frac{2}{3}{\sf R}^{2}-2{\sf R}_{\mu \nu }{\sf R}^{\mu
\nu })  \label{Efour}
\end{equation}%
where ${\sf R}_{\mu \nu }$and ${\sf R}$\ are the Ricci tensor and Ricci
scalar of \ the boundary spacetime. For the boundary metric of the bubble (%
\ref{Kerr5bb}), relation (\ref{Efour}) yields, 
\begin{equation}
E_{4}=\frac{a^{2}}{4\ell ^{4}}\{a^{2}X^{2}(3X^{2}+2)-\ell
^{2}(1+2X^{2})\}+O\left( \frac{1}{r^{2}}\right)  \label{Efourkerr}
\end{equation}%
Comparing (\ref{trace2}) with (\ref{Efourkerr}), we find 
\begin{equation}
\widetilde{T}_{\mu }^{\mu }=\beta E_{4}  \label{traceandanomaly4}
\end{equation}%
where $\beta =-\frac{\ell ^{3}}{2\pi G}=-\frac{N^{2}}{\pi ^{2}},$ in good
agreement with the prediction of AdS/CFT correspondence. In the above
calculation, we used the standard relation between the four-dimensional
Newton's constant and $N,$ which is\ given by $2N^{2}G=\pi \ell ^{3}$. The
bubble metric (\ref{Kerr5bb}),\ after the following coordinate
transformations: 
\begin{equation}
\begin{array}{c}
\text{\ }\Xi y^{2}\cosh ^{2}\Theta =(r^{2}-a^{2})(1+X^{2}) \\ 
y\sinh \Theta =rX \\ 
\Phi =\phi +a\chi /\ell ^{2}%
\end{array}
\label{cofcoord}
\end{equation}%
and a rescaling by the factor $\frac{\ell ^{2}}{y^{2}},$\ at large $y$,
takes the form 
\begin{equation}
ds_{b}^{\prime 2}=d\chi ^{2}+\ell ^{2}(-d\Theta ^{2}+\cosh ^{2}\Theta d\Phi
^{2}+\sinh ^{2}\Theta d\lambda ^{2})  \label{boundarymetr}
\end{equation}%
for $y\rightarrow \infty ,$\ which is a time dependent Einstein universe. \
We note all dependence on the rotational parameter disappears in boundary
metric when we go on the hypersurface of constant large $y.$\ This
observation is in agreement with the result on the boundary of \ Kerr-AdS
metric \cite{Hawk}. The metric (\ref{boundarymetr}) has a positive Ricci
scalar, which describes locally a $\left( 2+1\right) $-dimensional dS space
times a circle. Locally, the boundary metric (\ref{boundarymetr}) could be
written as 
\begin{equation}
d\widehat{s}_{b}^{2}=d\chi ^{2}+\ell ^{2}\{-d\Theta ^{2}+\cosh ^{2}\Theta
(d\alpha ^{2}+\sin ^{2}\alpha d\beta ^{2})\}  \label{boundarymetr2}
\end{equation}%
which by the Wick rotation $\Theta \rightarrow i\Omega $ changes to the
following static metric:%
\begin{equation}
d\widetilde{s}_{b}^{2}=d\chi ^{2}+\ell ^{2}\{d\Omega ^{2}+\cos ^{2}\Omega
(d\alpha ^{2}+\sin ^{2}\alpha d\beta ^{2})\}  \label{boundarymetr4}
\end{equation}%
and so the field theoretic trace anomaly calculation could be done on this
Euclidean space.

For the dual field theory on the boundary, since the metric (\ref%
{boundarymetr}), is conformally flat we can use the following relation for
the stress tensor \cite{Birrel}: 
\begin{equation}
\left\langle T_{\mu }^{\nu }\right\rangle =-\frac{1}{16\pi ^{2}}%
(A^{(1)}H_{\mu }^{\nu }+B^{(3)}H_{\mu }^{\nu })+\widetilde{T}_{\mu }^{\nu }
\label{s}
\end{equation}%
where $^{(1)}H_{\mu }^{\nu \text{ \ }}$and $^{(3)}H_{\mu }^{\nu }$\ are
conserved quantities constructed from the curvatures via 
\begin{equation}
\begin{array}{c}
^{(1)}H_{\mu \nu }=2R_{;\mu \nu }-2g_{\mu \nu }\square R-\frac{1}{2}g_{\mu
\nu }R^{2}+2RR_{\mu \nu } \\ 
^{(3)}H_{\mu \nu }=\frac{1}{12}R^{2}g_{\mu \nu }-R^{\rho \sigma }R_{\rho \mu
\sigma \nu }%
\end{array}
\label{H13}
\end{equation}%
and $\widetilde{T}_{\mu }^{\nu }$\ is the traceless part and the constants $%
A $\ and $B$\ are given by: $A=0$\ and $B\simeq \frac{\pi \ell ^{3}}{4G}$.
Using the boundary metric (\ref{boundarymetr}), we calculate the functions $%
^{(1)}H_{\mu }^{\nu \text{ \ }}$and $^{(3)}H_{\mu }^{\nu }$\ and find that
the first term of (\ref{s}) reproduces some parts of stress tensor (\ref%
{Kerr5bbstress}) which do not depend on bubble parameters $a$\ and $M.$

Similar results hold for bubbles obtained by double analytical continuation
of the two-rotation-parameter five- dimensional Kerr-AdS spacetime. The\
bubble metric after continuation is given by 
\begin{equation}
\begin{array}{c}
ds^{2}=\frac{\Delta _{r}}{\rho ^{2}}\{d\chi +\frac{\alpha }{\Xi _{a}}%
(1+X^{2})d\phi -\frac{b}{\Xi _{b}}X^{2}d\lambda \}^{2}+(1+X^{2})\frac{\Delta
_{X}}{\rho ^{2}}\{ad\chi -\frac{r^{2}-a^{2}}{\Xi _{a}}d\phi \}^{2}+\frac{%
\rho ^{2}}{\Delta _{r}}dr^{2} \\ 
+X^{2}\frac{\Delta _{X}}{\rho ^{2}}\{-bd\chi -\frac{r^{2}+b^{2}}{\Xi _{b}}%
d\lambda \}^{2}-\frac{\rho ^{2}}{(1+X^{2})\Delta _{X}}dX^{2}+\frac{1+\frac{%
r^{2}}{\ell ^{2}}}{r^{2}\rho ^{2}}\{abd\chi -b(1+X^{2})\frac{r^{2}-a^{2}}{%
\Xi _{a}}d\phi -aX^{2}\frac{r^{2}+b^{2}}{\Xi _{b}}d\lambda \}^{2}%
\end{array}
\label{Kerr5tworot}
\end{equation}%
where 
\begin{equation}
\begin{array}{c}
\Delta _{r}=(r^{2}-a^{2})(r^{2}+b^{2})(1+r^{2}/\ell ^{2})-2GM \\ 
\Delta _{X}=1-\frac{a^{2}X^{2}}{\ell ^{2}}-\frac{b^{2}(1+X^{2})}{\ell ^{2}}%
\text{ \ \ \ \ \ , \ \ \ \ \ }\Xi _{a}=1+\frac{a^{2}}{\ell ^{2}} \\ 
\rho ^{2}=r^{2}+a^{2}X^{2}+b^{2}(1+X^{2})\text{ \ \ , \ \ }\Xi _{b}=1-\frac{%
b^{2}}{\ell ^{2}}%
\end{array}
\label{delta4}
\end{equation}%
It also contains an ergoregion $\Delta _{X}<0$ ; hence (\ref{Kerr5bb})
describes a well defined bubble when $\left| X\right| <X_{cr}.$

The large-$r$ boundary diagonal stress-energy tensor components for the
metric (\ref{Kerr5tworot}) are

\begin{equation}
\begin{array}{c}
8\pi GT_{\phi }^{\phi }=\frac{1}{8\ell r^{4}}%
\{-7X^{4}(a^{2}+b^{2})^{2}+2X^{2}[-5a^{4}+\ell
^{2}(a^{2}+b^{2})-6b^{4}-11a^{2}b^{2}]+ \\ 
8GM\ell ^{2}-8a^{2}b^{2}-3a^{4}-4b^{4}+2a^{2}\ell ^{2}+\ell ^{4}\}+O\left( 
\frac{1}{r^{6}}\right) \\ 
8\pi GT_{X}^{X}=\frac{1}{8\ell r^{4}}%
\{-3X^{4}(a^{2}+b^{2})^{2}+2X^{2}[-a^{4}+\ell
^{2}(a^{2}+b^{2})-2b^{4}-3a^{2}b^{2}]+ \\ 
8GM\ell ^{2}+a^{4}+2a^{2}\ell ^{2}+\ell ^{4}\}+O\left( \frac{1}{r^{6}}\right)
\\ 
8\pi GT_{\chi }^{\chi }=\frac{1}{8\ell r^{4}}%
\{-7X^{4}(a^{2}+b^{2})^{2}-2X^{2}[a^{4}-5\ell
^{2}(a^{2}+b^{2})+6b^{4}+7a^{2}b^{2}]- \\ 
24GM\ell ^{2}+a^{4}-4b^{4}+2a^{2}\ell ^{2}+8b^{2}\ell ^{2}-3\ell
^{4}\}+O\left( \frac{1}{r^{6}}\right) \\ 
8\pi GT_{\lambda }^{\lambda }=\frac{1}{8\ell r^{4}}%
\{-7X^{4}(a^{2}+b^{2})^{2}+2X^{2}[-a^{4}+\ell
^{2}(a^{2}+b^{2})-2b^{4}-3a^{2}b^{2}]+ \\ 
8GM\ell ^{2}+a^{4}+2a^{2}\ell ^{2}+\ell ^{4}\}+O\left( \frac{1}{r^{6}}\right)%
\end{array}
\label{Kerr5bbtwoparstress}
\end{equation}%
and the resultant trace after rescaling is 
\begin{equation}
\widetilde{T}_{\mu }^{\mu }=\frac{1}{8\pi G\ell }%
\{-3(a^{2}+b^{2})^{2}X^{4}+2[-2b^{4}-3a^{2}b^{2}-a^{4}+(a^{2}+b^{2})\ell
^{2}]X^{2}+(a^{2}+b^{2})(\ell ^{2}-b^{2})\}+O\left( \frac{1}{r^{2}}\right)
\label{trace3}
\end{equation}%
The Euler density (\ref{Efour}), after a lengthy calculation is given by 
\begin{equation}
E_{4}=\frac{1}{4\ell ^{4}}%
\{3(a^{2}+b^{2})^{2}X^{4}-2[-2b^{4}-3a^{2}b^{2}-a^{4}+(a^{2}+b^{2})\ell
^{2}]X^{2}-(a^{2}+b^{2})(\ell ^{2}-b^{2})\}+O\left( \frac{1}{r^{2}}\right)
\label{Efourrkerrtworot}
\end{equation}

Once again, comparing our result (\ref{trace3}) with the Euler density of
the boundary theory (\ref{Efourrkerrtworot}), we find 
\begin{equation}
\widetilde{T}_{\mu }^{\mu }=\beta E_{4}  \label{traceandanomaly4tworot}
\end{equation}%
\ similar to the relation (\ref{traceandanomaly4}). The above result is in
good agreement with the AdS/CFT correspondence.

\section{Conclusions}

We have demonstrated that the AdS/CFT correspondence conjecture gives
reasonable results in time-dependent rotating bubble backgrounds: the
predicted relationship between the field-theoretic trace anomaly and the
asymptotic boundary stress-energy hold. This relationship is quite robust,
being satisfied not only for the conventional bubble metrics (\ref{Kerr7bb1}%
) and (\ref{Kerr7bb2}) in $d=7$, but also for the more complicated
time-dependent metrics (\ref{Kerr5bb}) and (\ref{Kerr5tworot}) \ with
ergoregions. \ This work also shows the usefulness of\ the counterterm
subtraction method applied to the time-dependent asymptotically AdS
spacetimes.{\large \ }An interesting problem would be to see what happens to
this prediction for the next-to-leading order terms. We leave this for
future consideration.

\medskip

{\Large Acknowledgments}

This work was supported by the Natural Sciences and Engineering Research
Council of Canada.\


\begin{thebibliography}{99}
\bibitem{AHA} O. Aharony, M. Fabinger, G. T. Horowitz and E. Silverstein,
JHEP {\bf 0207} (2002) 007.

\bibitem{Bir} D. Klemm, JHEP {\bf 9811} (1998) 019{\it ; }D. Birmingham and
M. Rinaldi, {\it Phys.Lett.} {\bf B544} (2002) 316{\it .}

\bibitem{bala} V. Balasubramanian and S. R. Ross, {\it Phys.Rev.} {\bf D66}
(2002) 086002{\it .}

\bibitem{CAI} R.-G. Cai, {\it Phys. Lett.} {\bf B544} (2002) 176{\it .}

\bibitem{MANN} A. M. Ghezelbash, R. B. Mann, JHEP {\bf 0209} (2002) 045.

\bibitem{dsCFT} A. Strominger, JHEP {\bf 0110} (2001) 034; E. Witten, {\it %
hep-th/0106109}; M. Li, JHEP {\bf 0204} (2002) 005; D. Klemm, {\it Nucl.Phys.%
} {\bf B625} (2002) 295; V. Balasubramanian, P. Horava and D. Minic, JHEP 
{\bf 0105} (2001) 043; A. Chamblin and N. D. Lambert, {\it Phys. Lett.} {\bf %
B508} (2001) 369; Y. Gao, {\it hep-th/0107067}; J. Bros, H. Epstein and U.
Moschella, {\it Phys. Rev.} {\bf D65} (2002) 084012; S. Nojiri and S. D.
Odintsov, JHEP {\bf 0112} (2001) 033; E. Halyo, \ {\it hep-th/0107169}; I.
Sachs and S. N. Solodukhin, {\it Phys. Rev.} {\bf D64} (2001) 124023; A. J.
Tolley and N. Turok, {\it hep-th/0108119}; S. Nojiri and S. D. Odintsov, 
{\it Phys. Lett.} {\bf B519 }(2001) 145; C. M. Hull, JHEP {\bf 0111} (2001)
012; S. Cacciatori and D. Klemm, {\it Class. Quant. Grav.} {\bf 19} (2002)
579; B. McInnes, {\it Nucl. Phys}. {\bf B627} (2002) 311; S. Nojiri and S.
D. Odintsov, {\it Phys. Lett. }{\bf B523} (2001) 165; A. Strominger, JHEP 
{\bf 0111} (2001) 049; V. Balasubramanian, J. de Boer and D. Minic, {\it %
Phys. Rev. }{\bf D65 }(2002) 123508; Y. S. Myung, {\it Mod. Phys. Lett.} 
{\bf A16} (2001) 2353; B. G. Carneiro da Cunha, {\it Phys. Rev.} {\bf D65}
(2002) 104025; R. Cai, Y. S. Myung, Y. Zhang, {\it Phys. Rev.} {\bf D65}
(2002) 084019; U. H. Danielsson, JHEP {\bf 0203} (2002) 020; S. Ogushi,\ 
{\it Mod. Phys. Lett.} {\bf A17} (2002) 51; A. C. Petkou and G. Siopsis, JHEP%
{\bf \ 0202 }(2002) 045; R. Cai, {\it Phys. Lett.} {\bf B525} (2002) 331; T.
Shiromizu, D. Ida and T. Torii, JHEP {\bf 11} (2001) 010; A. M. Ghezelbash
and R. B. Mann, JHEP {\bf 0201} (2002) 005; M. H. Dehghani, \ {\it Phys. Rev.%
} {\bf D65} (2002) 104003; A. M. Ghezelbash, D. Ida, R. B. Mann and T.
Shiromizu, {\it Phys. Lett.} {\bf B535} (2002) 315; M. Alishahiha and S.
Parvizi, JHEP {\bf 0210} (2002) 047; S. Nojiri and S. Odintsov, \ {\it Phys.
Lett. }{\bf B444} (1998) 92; S. Nojiri and S. Odintsov, {\it Phys. Lett. }%
{\bf B531} (2002) 143; R.-G. Cai and N. Ohta, {\it Phys. Rev. }{\bf D62}
(2000) 024006.

\bibitem{Hawk} S. W. Hawking, C. J. Hunter and M. M. Taylor-Robinson, {\it %
Phys. Rev.} {\bf D59} (1999) 064005.

\bibitem{awad1} A. M. Awad and C. V. Johnson, {\it Phys.Rev.} {\bf D61 }%
(2000) 084025.

\bibitem{awad2} A. M. Awad and C. V. Johnson, {\it Phys.Rev.} {\bf D63 }%
(2001) 124023.

\bibitem{Henning} M. Henningson and K. Skenderis, JHEP {\bf 9807} (1998) 023%
{\it .}

\bibitem{deharo} S. de Haro, K. Skenderis and S.N. Solodukhin, {\it Commun.
Math. Phys.} {\bf 217} (2001) 595.

\bibitem{cv} M. Cvetic, S. Nojiri and S.D. Odintsov, hep-th/0306031.

\bibitem{counterterm} S. Das and R.B. Mann, JHEP {\bf 0008} (2000) 033{\it ; 
}R.B. Mann, {\it Phys.Rev.} {\bf D60 }(1999) 104047{\it ; }R.B. Mann, {\it %
Phys.Rev.} {\bf D61 }(2000) 084013{\it ; }R. Emparan, C.V. Johnson and R.C.
Myers, {\it Phys.Rev.} {\bf D60 }(1999) 104001;{\it \ }V. Balasubramanian
and P. Kraus, {\it Commun. Math. Phys. {\bf 208} (1999) 413; }P. Kraus, F.
Larsen and R. Siebelink, {\it Nucl. Phys}. {\bf B563} (1999) 259.

\bibitem{Tseytlin} F. Bastianelli, S. Frolov and A. A. Tseytlin, JHEP {\bf %
0002} (2000) 013{\it .}

\bibitem{DW} B. S. De Witt, Dynamical theory of groups and fields, Gordon
and Breach, N.Y., 1965.

\bibitem{Birrel} N.D. Birrel and P.C. Davies, {\it Quantum Fields in Curved
Space, }Cambridge University Press, 1982{\it .}
\end{thebibliography}
\end{document}